\documentclass[useAMS,usenatbib]{mn2e}

\usepackage{graphicx}
\usepackage{verbatim}
\usepackage{fix2col}
\usepackage{amssymb}

\newcommand{\thetae}{\theta_{\rm E}}

\newcommand{\delxic}{\Delta \xi_{\rm c}}
\newcommand{\deletac}{\Delta \eta_{\rm c}}
\newcommand{\delxiw}{\Delta \xi_{\rm w}}
\newcommand{\deletaw}{\Delta \eta_{\rm w}}

\title[Planetary Caustic Perturbation]{Properties of the
Planetary Caustic Perturbation}

\author[S.-J. Chung and C.-U. Lee]
{Sun-Ju Chung$^1$\thanks{E-mail; sjchung@kasi.re.kr (SJC); leecu@kasi.re.kr
(CUL)} and Chung-Uk Lee$^{1\star}$\\
$^{1}$ Korea Astronomy and Space Science Institute, Hwaam-Dong,
Yuseong-Gu, Daejeon 305-348, Korea \\
}

\begin{document}
\date{Accepted......Received......;in original form 2010 25 June}

\pagerange{\pageref{firstpage}--\pageref{lastpage}} \pubyear{2010}
\maketitle

\label{firstpage}

\begin{abstract}
Just two of 10 extrasolar planets found by microlensing have been detected by the planetary caustic despite the higher probability of planet detection relative to the central caustic which has been responsible for four extrasolar planet detections.
This is because the perturbations induced by the planetary caustic are unpredictable, thus making it difficult to carry out strategic observations.
However, if future high-cadence monitoring surveys are conducted, the majority of planetary caustic events including the events by free-floating planets and wide-separation planets would be detected.
Hence, understanding the planetary caustic perturbations becomes important.
In this paper, we investigate in detail the pattern of the planetary caustic perturbations.
From this study, we find three properties of the planetary caustic perturbations.
First, planetary systems with the same star-planet separation ($s$) basically produce perturbations of constant strength regardless of the planet/star mass ratio ($q$), but the duration of each perturbation scales with $\sqrt{q}$.
Second, close planetary systems with the same separation produce essentially the same negative perturbations between two triangular-shaped caustics regardless of $q$, but the duration of the perturbations scales with $\sqrt{q}$.
Third, the positive perturbations for planetary systems with the same mass ratio become stronger as the caustic shrinks with the increasing $|\log s|$, while the negative perturbations become weaker.
We estimate the degeneracy in the determination of $q$ that occurs in planetary caustic events.
From this, we find that the mass ratio can be more precisely determined as $q$ increases and $|\log s|$ decreases.
We also find that the degeneracy range of events for which the source star passes close to the planetary caustic is usually very narrow, and thus it would not significantly affect the determination of $q$.
\end{abstract}


\section{INTRODUCTION}

The microlensing signal of a planet is a short-duration perturbation on the smooth standard light curve of the primary-induced lensing event occurring on a background source star.
The planetary perturbation is induced by the central and planetary caustics, which are typically separated from each other.
The central caustic is always formed close to a host star and thus the perturbation induced by the central caustic occurs near the peak of the lensing light curve.
In events induced by the central caustic, there exists a $s \leftrightarrow 1/s$ degeneracy, where $s$ represents the projected star-planet separation normalized to the Einstein radius of the lens system (Griest \& Safizadeh 1998; Dominik 1999).
The degeneracy arises due to the similarity of the size and shape of the central caustics for $s$ and $1/s$.
Since the difference of the size and shape of the two central caustics increases as the planet/star mass ratio increases, the degeneracy can be broken for events involving giant planets \citep{chung05}.
High-magnification events for which the source star passes close to the host star are very sensitive for the detection of a planet due to the central caustic \citep{griest98}.
Thus, current microlensing follow-up observations ($\mu$FUN: Dong et al. 2006; PLANET: Albrow et al. 2001; RoboNet: Burgdorf et al. 2007), which intensively monitor events alerted by microlensing survey observations (OGLE: Udalski 2003; MOA: Bond et al. 2002), are biased toward high-magnification events.
Due to the reason, four of 10 extrasolar planets found by microlensing (Udalski et al. 2005; Gaudi et al. 2008; Bennett et al. 2008; Dong et al. 2009) were detected by the central caustic.

On the other hand, the planetary caustic is formed away from the host star and thus the perturbation induced by the planetary caustic can occur at any part of the lensing light curve.
The planetary caustic is much bigger than the central caustic, and thus the probability of detecting a planet by the planetary caustic is much higher than by the central caustic.
In spite of the advantage of high detection efficiency, only two of 10 microlensing extrasolar planets (Beaulieu et al. 2006; Sumi et al. 2010) were detected by the planetary caustic.
This is because it is hard to carry out strategic observations due to the unpredictable nature of the planetary caustic perturbation.
However, if future ground- and space-based surveys with high-cadence monitoring, such as KMTNet (Korean Microlensing Telescope Network; B. Park 2010, private communication) and \emph{MPF} (\emph{Microlensing Planet Finder}; D. Bennett 2004), are conducted, the majority of the planetary caustic events including the events by free-floating planets and wide-separation planets would be detected.
Hence, understanding the planetary caustic perturbations becomes important.
In addition, for the planetary caustic events, there is a degeneracy in the determination of the planet/star mass ratio (Gould \& Loeb 1992; Gaudi \& Gould 1997), which is derived from the duration of the planetary caustic perturbation relative to the total duration of the event.
To find out how much the degeneracy affects the determination of the mass ratio, the estimation of the degeneracy is needed.

In this paper, we investigate in detail the pattern of the planetary caustic perturbations and estimate the degeneracy in the determination of the planet/star mass ratio in planetary caustic events.
This paper is organized as follows. In Section 2, we briefly describe the planetary lensing.
In Section 3, we investigate perturbation patterns of the planetary caustics for various planetary systems.
We also estimate the degeneracy in the mass ratio in the planetary caustic events.
We summarize the results in Section 4.

\section{PLANETARY LENSING}

Planetary lensing is described by the special case of binary lensing with very low mass ratio and the lensing behavior is usually well described by that of a single lensing of the host star for most of the event duration.
In this case, the lens equation \citep{witt90} is expressed as
\begin{equation}
\zeta = z - {1\over{1+q}} \left({ 1\over{\bar{z}}} + {q\over{\bar{z} - \bar{z}_p}}\right),
\end{equation}
where $\zeta = \xi + i\eta$ and $z = x + iy$ represent the complex notations of the source and image positions, respectively, $\bar{z}$ denotes the complex conjugate of $z$,  $z_p$ is the position of the planet, and $q$ is the planet/star mass ratio.
Here the position of the star is at the center and all lengths are normalized to the Einstein ring radius of the total mass of the lens system, $\thetae$.

In planetary lensing, the shape and number of the planetary caustics depend on the star-planet separation unlike the central caustics, which are always single and arrow-shaped.
For close planets ($s < 1$), there are two triangular-shaped caustics with three cusps.
The two caustics are symmetrically displaced perpendicular to the star-planet axis and located at the opposite side to the planet.
The horizontal and vertical widths of the caustic, defined as the separations between the on- and off-axis cusps \citep{han06}, are expressed respectively as
\begin{equation}
\delxic \sim {{3\sqrt{3} \over 4} \sqrt{q}s^{3}},\\
\deletac \sim \sqrt{q}s^{3}.\
\end{equation}
The distance between the two caustics is represented by
\begin{equation}
y = {4\sqrt{q}(1-s^{2})^{1/2} \over s} .
\end{equation}
Events for which the source passes between the two caustics produce negative perturbations on the lensing light curves (Gaudi \& Gould  1997; Han \& Chang 2003).

For wide planets ($s > 1$), there is a single asteroid-shaped caustic with four cusps, where the horizontal width along the star-planet axis is generally longer than the vertical width.
The horizontal and vertical widths of the caustic \citep{han06} are expressed respectively as
\begin{equation}
\delxiw \sim {4\sqrt{q}\over s^{2}}\left (1 + {1 \over 2s^{2}}\right ),\\
\deletaw \sim {4\sqrt{q} \over s^{2}}\left (1 - {1 \over 2s^{2}}\right ) .
\end{equation}
The caustic for $s \gg 1$ becomes a symmetric diamond-shaped caustic, while for $s \rightarrow 1$ it becomes more asymmetric.
The caustic is located at the same side as the planet.
Events induced by the wide planetary caustic produce positive perturbations on the lensing light curves (Gaudi \& Gould  1997; Han \& Chang 2003).
The position of the close and wide caustics is always located at $s - 1/s$ from the host star.

The lensing behavior of the planetary caustic can be described by Chang-Refsdal lensing, which represents single lensing superposed on a uniform background shear (Chang \& Refsdal 1979; Gould \& Loeb 1992; Gaudi \& Gould 1997). In this case, the single lensing is produced by the planet itself and the shear is induced by the host star.
The lens equation of the Chang-Refsdal lensing is represented by
\begin{equation}
\zeta = z - {1\over{\bar{z}}} + \gamma \bar{z} \ ;\ \ \ \gamma = {1\over
s^2},
\end{equation}
where $\gamma$ is the shear.

\begin{figure}
\includegraphics[width=84mm]{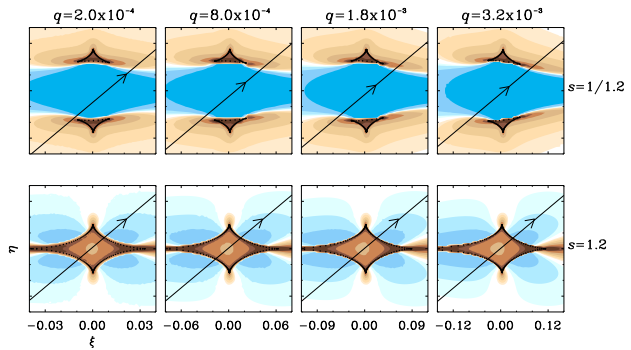}
\caption{
Magnification excess maps of close and wide planetary systems with various mass ratios. The separations of close and wide systems are $s = 1/1.2$ and $s = 1.2$, respectively. The coordinates ($\xi$, $\eta$) represent the axes that are parallel with and normal to the star-planet axis and are centered at the
planetary caustic center. The color changes into darker scales when the excess is $|\epsilon| = 2\%,\ 4\%,\ 8\%,\ 16\%,\ 32\%,\ 64\%,\ \rm {and},\ 96\%$, respectively. The straight lines with arrows represent the source trajectories for which light
curves of the resulting events are presented in Figure 3.
}
\end{figure}

\section{PLANETARY CAUSTIC PERTURBATION}

To investigate the perturbation pattern of the planetary caustic, we produce magnification excess maps of various planetary systems.
The magnification excess is defined by
\begin{equation}
\epsilon = {A-A_{0}\over A_{0}},
\end{equation}
where $A$ and $A_0$ are the planetary and single lensing magnifications, respectively.

Figures 1 and 2 show the magnification excess maps of planetary systems with various mass ratios and separations.
In each map, the regions with blue and brown-tone colors represent the areas where the excess is negative and positive, respectively.
The color changes into darker scales when the excess is $|\epsilon| = 2\%,\ 4\%,\ 8\%,\ 16\%,\ 32\%,\ 64\%,\ \rm {and},\ 96\%$, respectively.
Since the excess represents the perturbation induced by the planetary caustic, increasing the excess represents that the strength of the perturbation becomes strong.
In Figure 1, the top and second rows show close and wide planetary systems with $s = 1/1.2$ and $s = 1.2$, where each column corresponds to a different mass ratio.
In Figure 2, the top and second rows show close and wide planetary systems with different separations, where each panel has a common mass ratio of $q = 10^{-3}$.
Considering typical microlensing events toward Galactic bulge, we assume that $M_{\rm L} = 0.3\ M_{\odot}$, $D_{\rm L} = 6\ {\rm kpc}$, $D_{\rm S} = 8\ {\rm kpc}$, and the source is a main-sequence star with a radius of $R_{\star} = 1.0\ R_{\odot}$.
Then, the corresponding angular Einstein radius is $\thetae = 0.32\ \rm{mas}$, and the source radius normalized to the Einstein radius is $\rho_{\star} = 1.8\times10^{-3}$.

\begin{figure}
\includegraphics[width=84mm]{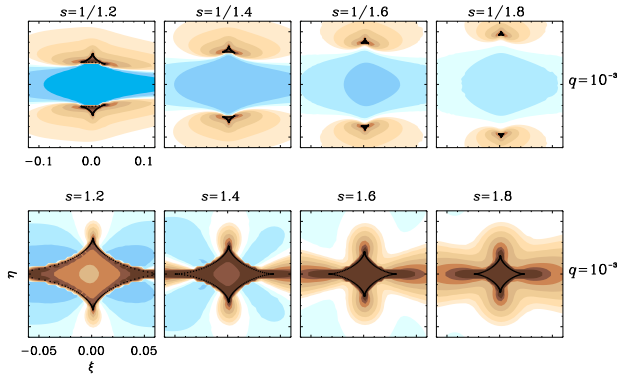}
\caption{
Magnification excess maps of close and wide planetary systems with various
separations.
All the systems have a common mass ratio of $q = 10^{-3}$.
}
\end{figure}

From the analysis of the maps, we find the following properties of the planetary caustic perturbations.
\begin{enumerate}
\item
Planetary systems with the same separation basically produce perturbations of constant strength regardless of $q$, but the duration of each perturbation scales with $\sqrt{q}$.
\item
Close planetary systems with the same separation produce essentially the same negative perturbations between two triangular-shaped caustics regardless of $q$, but the duration of the perturbations scales with $\sqrt{q}$.
\item
For planetary systems with the same mass ratio, the positive perturbations become stronger as the caustic shrinks with the increase (for $s > 1$) or the decrease (for $s < 1$) of $s$.
On the other hand, the negative perturbations become weaker as the caustic shrinks.
\end{enumerate}
The fundamental reason for three properties above is that the lens equation of the binary lens near the location of the planet is approximated as that of a Chang-Refsdal lens which is a point mass on a uniform background shear (Gould \& Loeb 1992; Gaudi \& Gould  1997).
In this case, the point mass is the planet and the shear is induced by the host star.
Thus, if the separation is the same, then the shear, $\gamma \sim 1/s^2$, which causes the perturbation, is also the same.
The three properties imply that the planetary caustic perturbation depends only the star-planet separation.
This fact has been discussed in \citet{gould92}, \citet{gaudi97}, and \citet{dominik99}.
However, they did not summarize in detail the properties of the planetary caustic perturbation such as we have done.
In addition, the third property related to the positive perturbation has never been reported.

For planetary systems for which the source is similar to or bigger than the planetary caustic, the perturbations around the caustic are significantly washed out by the increase of finite-source effects and thus become weak.
However, since the magnification patterns of the planetary caustics at the same separation are basically constant, as long as the ratio of the source size to the caustic size is the same, the first and second properties above are always maintained even in planetary systems with significant finite-source effects \citep{gould97}.
The third property is also maintained in planetary systems with significant finite-source effects, because the positive perturbations becomes stronger as the caustic shrinks.
Therefore, the three properties above are always maintained in planetary systems with significant finite-source effects, as long as the ratio of the source size to the caustic size is the same.

\begin{figure}
\includegraphics[width=84mm]{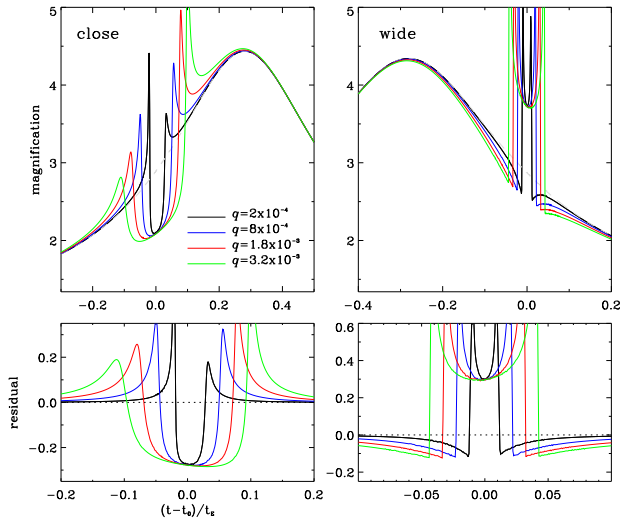}
\caption{
Light curves for the source trajectories presented in Figure 1. In the upper panel, solid and dashed curves represent the light curves of the planetary and single lensing events, respectively. The lower panel shows the residuals from the single lensing magnification. In the lower panel, the horizontal line indicates the magnification excess of $|\epsilon| = 0.0$.
}
\end{figure}

\begin{table}
\caption{Degeneracy Range for Close Planetary Lensing Systems.\label{tbl-one}}
\begin{tabular}{lccc}
\hline\hline
 & & \multicolumn{2}{c}{Degeneracy Range ($\Delta{q}/q_{0}$)}\\ &&\multicolumn{2}{c}{(\%)} \\ \cline{3-4} \\
(a) & Planet/Star Mass Ratio ($q_0$)  & $\rho_\star$ = 0.0018 & $\rho_\star$ = 0.0054\\
\hline
& $10^{-5}$               & 14.7 & 40.0 \\
& $5\times10^{-5}$   & 10.1 & 17.8 \\
& $10^{-4}$               & 7.3   & 12.7 \\
& $5\times10^{-4}$   & 5.4   & 8.9 \\
& $10^{-3}$               & 4.0   & 6.2 \\
& $5\times10^{-3}$   & 1.4   & 2.3 \\
\hline
(b) & Planet-Star Separation ($s$)  &    &\\
\hline
& $1/1.3$   & 16.7  & 18.2 \\
& $1/1.5$   & 69.6  & 69.6 \\
& $1/1.7$   & 199.0   & 185.0  \\
& $1/1.9$   & 877.0   & 881.0 \\
\hline
(c) & Source Trajectory Angle ($\alpha$)  &    &\\
\hline
&$20^\circ$   & 20.8  & 22.2 \\
&$40^\circ$   & 4.0  & 6.4 \\
&$60^\circ$   & 0.5   & 0.8  \\
&$80^\circ$   & 0.4   & 1.4 \\
\hline
\end{tabular}

\medskip
Notes. Degeneracy range with $|\delta| \le 5\%$.
(a)  The planet-star separation is $s=1/1.2$.  (b) The planet/star mass ratio is $q_{0} = 10^{-3}$.  The source trajectory angle from the planet-star axis, $\alpha$,  for (a) and (b) is 40$^{\circ}$. (c) The mass ratio and separation are $q_{0} = 10^{-3}$ and $s = 1/1.2$. Here the adopted source radii are $R_\star = 1$ and $3\ R_\odot$ for the main-sequence and turnoff stars, respectively. The normalized source radii are determined by an Einstein radius of $\thetae = 0.32\ \rm{mas}$ that corresponds to the value of the typical Galactic bulge event with $M_{\rm L} = 0.3\ M_{\odot}$, $D_{\rm L} = 6\ {\rm kpc}$, and
$D_{\rm S} = 8\ {\rm kpc}$.
\end{table}

Figure 3 shows the light curves resulting from the source trajectories presented in Figure 1.
Figure 3 describes well the first and second properties of the planetary caustic perturbations.
Close planetary systems with the same separation produce negative perturbations with a same depth on the light curves, while wide planetary systems produce positive perturbations with a same bottom of U-shape.
The widths of the negative and positive perturbations increase as $q$ increases, because the distance between two close planetary caustics and size of the wide planetary caustic increase with the increase of $q$.
Since the width of the perturbation is used to determine the mass ratio (Gould \& Loeb 1992; Gaudi \& Gould  1997), the first and second properties above could cause a degeneracy in the determination of the mass ratio.
To find out how much the degeneracy affects the determination of $q$, we estimate the degeneracy.
Tables 1 and 2 show the degeneracy ranges of close and wide planetary lensing systems, respectively.
In each table, the degeneracy range is presented for various mass ratios (a), separations (b), and source trajectory angles (c) with two kinds of source radii.
The degeneracy range represents the range with $|\delta| \le 5\%$, where $\delta$ is defined by
\begin{equation}
\delta = {A_{\rm p} - A_{\rm p,0}\over{A_{\rm p,0}}},
\end{equation}
where $A_{\rm p}$ and $A_{\rm p,0}$ are the magnifications of the planetary lensing events with trial and fiducial mass ratios, respectively.
In the tables, $q_0$ is the fiducial mass ratio and $\Delta q$ is the difference of maximum and minimum trial mass ratios with $|\delta| \le 5\%$.
As shown in the two tables, the degeneracy range decreases with the increase of $q$ and decreases of $|\log s|$ and the finite-source effect.
It implies that we can determine $q$ more precisely as $q$ increases and $|\log s|$ decreases.
For close planetary systems, the degeneracy range dramatically increases as the separation decreases.
This is because the gap between two close caustics become wider and the perturbation becomes weak as the separation decreases, and thus the source passes only weak perturbation regions with $|\epsilon| \le 5\%$.
However, as shown in the case of (c) of Table 1, the degeneracy range considerably decreases as the trajectory angle increases, i.e., the source approaches the caustic.
On the other hand, for wide planetary systems, the degeneracy range is for caustic-crossing events.
The degeneracy range increases as the trajectory angle increases.
This is because the region close to the planet-star axis produces strong perturbations relative to other regions (see the bottom panel in Figure 1).
Therefore, the degeneracy range of events for which the source passes close to the caustic is usually very narrow, and thus it would not significantly affect the determination of $q$.
The degeneracy can also be easily resolved by the full coverage of the perturbation.
However, if the perturbation is covered by only a few data, it is difficult to determine the exact mass ratio.

\begin{table}
\caption{Degeneracy Range for Wide Planetary Lensing Systems.\label{tbl-one}}
\begin{tabular}{lccc}
\hline\hline
& & \multicolumn{2}{c}{Degeneracy Range ($\Delta{q}/q_{0}$)}\\ &&\multicolumn{2}{c}{(\%)} \\ \cline{3-4} \\
(a) & Planet/Star Mass Ratio ($q_0$) & $\rho_\star$ = 0.0018 & $\rho_\star$ = 0.0054\\
\hline
& $10^{-5}$               & 47.0 & 82.0 \\
& $5\times10^{-5}$   & 13.8 & 55.4\\
& $10^{-4}$               & 7.3   & 31.0 \\
& $5\times10^{-4}$   & 1.9   & 8.8 \\
& $10^{-3}$               & 0.9   & 5.8 \\
& $5\times10^{-3}$   & 0.3   & 2.0 \\
\hline
(b) & Planet-Star Separation ($s$)  &    &\\
\hline
& $1.3$   & 1.0  & 4.9 \\
& $1.5$   & 0.9  & 3.9 \\
& $1.7$   & 2.3   & 4.3  \\
& $1.9$   & 3.4   & 5.9 \\
\hline
(c) & Source Trajectory Angle ($\alpha$)  &    &\\
\hline
&$20^\circ$   & 1.0  & 5.1 \\
&$40^\circ$   &  0.9 & 6.7 \\
&$60^\circ$   & 1.5   & 8.5  \\
&$80^\circ$   & 1.9   & 10.5 \\
\hline
\end{tabular}

\medskip
Note. Except for $s = 1.2$ in (a) and (c), other parameters are the same with those of Table 1.
\end{table}

\section{CONCLUSION}

We have investigated the pattern of planetary caustic perturbations.
From this, we found three properties of the planetary caustic perturbations.
First, planetary systems with the same star-planet separation basically produce the perturbations with constant strength without regard to $q$, but the duration of each perturbation is proportional to $\sqrt{q}$.
Second, close planetary systems with the same separation produce essentially the same negative perturbations between two triangular-shaped caustics regardless of $q$, but the duration of the perturbation scales with $\sqrt{q}$.
Third, the positive perturbations for planetary systems with the same mass ratio become stronger as the caustic shrinks with the increase ($s > 1$) or the decrease ($s < 1$) of $s$, while the negative perturbations become weaker.
We have estimated the degeneracy in the determination of $q$ that occurs in planetary caustic events.
From this, we found that $q$ can be more precisely determined as $q$ increases and $|\log s|$ decreases.
We also found that the degeneracy range of events for which the source passes close to the caustic is usually very narrow, and thus it would not significantly affect the determination of $q$.

\section*{Acknowledgments}
We would like to thank A. Gould for making helpful comments and suggestions.

\label{lastpage}


\begin{thebibliography}{99}

\bibitem[\protect\citeauthoryear{Albrow et al.}{2001}]{albrow01}
Albrow, M. D., et al. 2001, ApJ, 556, L113

\bibitem[\protect\citeauthoryear{Beaulieu et al.}{2006}]{beaulieu06}
Beaulieu, J.-P., et al. 2006, Nature, 439, 437

\bibitem[\protect\citeauthoryear{Bennett et al.}{2008}]{bennett08}
Bennett, D. P., et al. 2008, ApJ, 684, 663

\bibitem[\protect\citeauthoryear{Bond et al.}{2002}]{bond02}
Bond, I. A., et al. 2002, MNRAS, 331, L19

\bibitem[\protect\citeauthoryear{Burgdorf et al.}{2007}]{burgdorf07}
Burgdorf, M. J., et al. 2007, Planet. Space Sci., 55, 582

\bibitem[\protect\citeauthoryear{Chang \& Refsdal}{1979}]{chang79}
Chang, K., Refsdal, S., 1979, Nature, 282, 561

\bibitem[\protect\citeauthoryear{Chung et al.}{2005}]{chung05}
Chung, S.-J., et al. 2005, ApJ, 630, 535

\bibitem[\protect\citeauthoryear{Dominik}{1999}]{dominik99}
Dominik, M. 1999, A\&A, 349, 108

\bibitem[\protect\citeauthoryear{Dong et al.}{2006}]{dong06}
Dong, S., et al. 2006, ApJ, 642, 842

\bibitem[\protect\citeauthoryear{Dong et al.}{2009}]{dong09}
Dong, S., et al. 2009, ApJ, 698, 1826

\bibitem[\protect\citeauthoryear{Gaudi \& Gould}{1997}]{gaudi97}
Gaudi, B. S., Gould, A. 1997, ApJ, 486, 85

\bibitem[\protect\citeauthoryear{Gaudi et al.}{2008}]{gaudi08}
Gaudi, B. S., et al. 2008, Science, 319, 917

\bibitem[\protect\citeauthoryear{Gould \& Loeb}{1992}]{gould92}
Gould, A., Loeb, A. 1992, ApJ, 396, 104

\bibitem[\protect\citeauthoryear{Gould \& Gaucherel}{1997}]{gould97}
Gould, A., Gaucherel, C. 1997, ApJ, 477, 580

\bibitem[\protect\citeauthoryear{Griest \& Safizadeh}{1998}]{griest98}
Griest, K., Safizadeh, N. 1998, ApJ, 500, 37

\bibitem[\protect\citeauthoryear{Han}{2006}]{han06}
Han, C. 2006, ApJ, 638, 1080

\bibitem[\protect\citeauthoryear{Han \& chang}{2003}]{han03a}
Han, C., Chang, K. 2003, ApJ, 597, 1070

\bibitem[\protect\citeauthoryear{Sumi et al.}{2010}]{sumi10}
Sumi, T., et al. 2010, ApJ, 710, 1641

\bibitem[\protect\citeauthoryear{Udalski}{2003}]{udalski03}
Udalski, A. 2003, Acta Astron., 53, 291

\bibitem[\protect\citeauthoryear{Udalski et al.}{2005}]{udalski05}
Udalski, A., et al. 2005, ApJ, 628, L109

\bibitem[\protect\citeauthoryear{Witt}{1990}]{witt90}
Witt, H. J. 1990, A\&A, 236, 311

\end{thebibliography}
\end{document}